\documentclass[10pt,letterpaper]{article}
\usepackage[top=0.85in,left=2.75in,footskip=0.75in]{geometry}

\usepackage{changepage}

\usepackage[utf8]{inputenc}

\usepackage{textcomp,marvosym}

\usepackage{fixltx2e}

\usepackage{amsmath,amssymb}

\usepackage{cite}

\usepackage{nameref,hyperref}

\usepackage[right]{lineno}

\usepackage{microtype}
\DisableLigatures[f]{encoding = *, family = * }

\usepackage{rotating}

\raggedright
\usepackage[aboveskip=1pt,labelfont=bf,labelsep=period,justification=raggedright,singlelinecheck=off]{caption}

\makeatletter
\renewcommand{\@biblabel}[1]{\quad#1.}
\makeatother

\date{}

\usepackage{lastpage,fancyhdr,graphicx}
\usepackage{epstopdf}
\fancyheadoffset[L]{2.25in}
\fancyfootoffset[L]{2.25in}
\usepackage[utf8]{inputenc}
\usepackage[english]{babel}
\usepackage{graphicx}
\usepackage{listings}
\usepackage{subcaption}

\begin{document}
\vspace*{0.35in}

\begin{flushleft}
{\Large
\textbf\newline{No cell left behind: automated physics-based tracking of {\em every} cell in a dense and growing colony}
}
\newline
\\
Huy Pham,
Emile Ramez Shehada,
Shawna Stahlheber,
Wayne B. Hayes\ddag*
\\
\bigskip
Department of Computer Science, University of California, Irvine, CA, 92697-3435, USA
\\
\bigskip

%
%

\ddag Senior author
* whayes@uci.edu

\end{flushleft}
\section*{Abstract} 

A human watching a video of closely-packed cells can generally identify {\em every} individual cell, regardless of density and noise, but most currently-available cell-tracking software cannot. This is because the human brain automatically builds a physical model of the scene as it progresses, allowing it to readily distinguish cells from noise and not be unduly confused by overlapping cells. Here we introduce software that uses physical rules to create a simulation of the activity in a cell video, synchronizing itself with the video as the activity progresses. Because our simulation includes every individual cell, we are trivially able to track all cell movement, growth, and divisions. Our method is also particularly robust to noise without requiring any substantial image processing. We demonstrate the effectiveness of this method by tracking the motion and lineage tree of a densely-packed colony of cells that grows from 4 to more than 200 individuals.



\section{Introduction and Motivation}

If you, as a human, watch a video of people walking around a park, you don't get confused when somebody momentarily disappears behind a tree and then reappears. This is because you have a model of the world in your head that says people who disappear still exist even though you can't see them, and that people can't just appear out of nowhere. This model also allows you to correctly interpret noisy observations by allowing you to safely dismiss noise that cannot possibly represent anything real. Unfortunately, existing cell tracking software largely isn't that smart; if cells are too closely packed or too numerous or the video is too noisy, individuals will get lost between frames, with the cumulative loss rate meaning that no individual can be traced through more than a few frames, making it impossible to track individuals; consequently, no cell's lineage is known with any certainty and even remotely correct family trees are impossible to build. Furthermore, detailed motion of individual cells cannot be studied using existing software, unless {\em immense} human time is spent curating video frames.

The solution is to have a system that understands what can plausibly happen to cells between video frames: they can move, rotate, and change shape a bit; they can split; they can even vanish from view, leave the frame, or appear on the edge of a frame. However, they cannot vanish from existence, and momentary noise should never be interpreted as a new cell that appears out of nowhere.\footnote{For now we ignore the case of cells moving in 3D that drift in-and-out of a focal plane, which {\em can} look like cells (dis)appearing out of nowhere. Such a case can easily be programmed into our system.}

Accordingly, in this paper we introduce {\it Cell Universe}, a simple proof-of-concept algorithm that tracks {\em every} individual in a population of bacterial cells as they move, grow, and reproduce on a 2-dimensional glass plate, packed so closely that they often touch or overlap. We emphasize that {\em every single member} of the colony is tracked, without fail, without losing anybody and without adding spurious members, as the population grows from 4 cells to more than 200; errors start to occur only when the video image degrades to the point that even a human would have difficulty figuring out what is happening. We do this by building a physical {\em model} of the colony and all its members in a simulation (the ``universe''), initialized in the first frame. The universe is then duplicated into an ensemble of nearly identical universes, each member of which is advanced stochastically via physics-based rules to plausible futures at the next video frame. Then, all universes are discarded except those that most closely correspond to the actual next frame of the video. Our method is {\em extremely} robust to image noise, and accounts for the temporary disappearance of cells from {\em view}---but not from the {\em universe}---so no individual is ever lost. The process of building ensembles of universes each closely representing reality is elsewhere known as ``ensemble simulation;'' pruning the possibilities based on a continually updated stream of real-world data is called ``data assimilation.'' To our knowledge, this is the first time they have been used together in the cell-tracking arena.

This method allows us to reliably track the detailed motion and life cycle of {\em every} individual in a densely populated and growing two-dimensional bacterial colony, allowing precise study of bacterial colonies; we can robustly build family trees, gather statistics on motion, individual growth, local and global reproduction rates, and perform statistical analyses of the entire culture with high accuracy and detail. For example, we observe that as the colony matures, individuals near the edge of the culture reproduce slightly more quickly than those near the center, presumably due to depletion of resources near the center. Such a subtle effect would be impossible to detect without the accurate tracking described herein, since one needs the history provided by a reliable family tree to discern that the cells on the periphery are exactly one generation ahead of the rest of the colony.
The method is readily generalizable to three dimensions and to more general groups of cells than bacteria.



\section{Previous Work}

Two broad categories of cell tracking methods are described in the existing literature: tracking by detection and tracking by model evolution \cite{li_cell_2008, maska_benchmark_2014}. Tracking by detection is a two-stage method. The first stage involves processing each frame of a cell video to identify individual cells, segmenting on the basis of gradient, intensity, textures, etc. The second stage involves applying an optimization strategy to determine cell correspondence from frame to frame \cite{meijering_tracking_2009}. 

There are numerous tracking-by-detection methods detailed in the literature. Most algorithms use some variation of intensity of the image. For example, Li et al use an adaptive thresholding of intensity, optimized at each pixel via integer programming into a binary cell detection at that pixel \cite{li_multiple_2010}. Seeded watershed algorithms primarily use intensity and its gradient to isolate probable locations of cells \cite{al-kofahi_automated_2006}. Gradient-based edge-detection attempts to isolate the boundary between a cell and its surroundings \cite{zhang_collective_2010}. Shift-invariant wavelet frame transformations can isolate cells by optimizing coupled minimum-cost flow tracking \cite{padfield_coupled_2011}. Most open-source, readily-available software falls within this paradigm, including CellTrack \cite{sacan_celltrack:_2008}, CellCounter \cite{li_cellcounter:_2014}, and most prominently, CellProfiler \cite{carpenter_cellprofiler:_2006, lamprecht_cellprofiler:_2007}. 

Tracking-by-detection methods can be computationally inexpensive, but many of them require the user to estimate the gating threshold \cite{al-kofahi_automated_2006, jaqaman_robust_2008}. Some advancements in the tracking-by-detection paradigm directed at remedying these problems include SAMTRA \cite{kan_automated_2011}, which attempts to automatically determine gating thresholds, and Lineage Mapper \cite{chalfoun_lineage_2016}, which uses segmented masks to bypass dependency on any particular segmentation method. Some tracking-by-detection methods do not require the user to estimate the gating threshold (for example, \cite{padfield_coupled_2011}), but these methods typically have less robust noise-handling techniques \cite{meijering_tracking_2009, kan_automated_2011}. 

In contrast, tracking by model evolution involves simultaneously segmenting and tracking cells in each frame of a cell video; the results of each frame are used to initialize the analysis of the following frame. This paradigm more easily accommodates morphological features of cells or user information about cell behavior. For example, models may be designed to allow cell division but prevent cell fusion \cite{zhang_tracking_2004}. Most model evolution methods evolve the contours of the cells \cite{dzyubachyk_automated_2010, maska_segmentation_2013, baker_automated_2014}. Other approaches rely on topology-constrained level set methods \cite{nath_cell_2006, li_cell_2008}.

Although our method falls squarely within the model evolution paradigm, we do not rely on contour tracking or similar methods. Instead, we make use of ensemble simulation. Ensemble simulation is a method of tracking and understanding real-world systems by building a realistic computer simulation of the system, and then having the simulation initialized with the state of the real system. The simulation is then run forward in simulated time; if the simulation is sufficiently realistic, then it will follow approximately the same path as the real system for some nontrivial duration of time. After some time, we stop the simulation, wait for the real system to catch up if necessary, and then ``pull'' the state of the simulation towards the real system in order to re-synchronize the two. In fact, since measurements of a real system have observational error, we actually initialize {\em multiple} states of the simulated system, all within the observational uncertainty of the real system. This is the {\it ensemble} of simulations. They all run forward in time, and when the real system ``catches up'', the process of choosing which ensembles best approximate the real system, and re-synchronizing the real and simulated systems is called {\it data assimilation}. Ensemble simulation is heavily used in weather prediction, where it has a long and rich history. It has been used in a diverse array of applications, including global \cite{whitaker_ensemble_2008} and regional \cite{wang_central_2011} weather forecasting, flood forecasting \cite{demeritt_ensemble_2007}, and carbon dioxide emissions simulations \cite{dai_ensemble_2001}. 
This method is easily adapted to tracking cells. Once the first frame is initialized, we use an entirely physics-based simulation to {\em predict} many possible locations of each cell in the next frame, and then use the next frame to {\em pick} which of the predictions is actually realized. Cells can never be lost to noise or fuse with other cells because disappearances and mergers are simply not physically possible in our model. As a result, our method is extremely resistant to noise. 

Our method is somewhat reminiscent of the method of multi-hypothesis tracking described by Jaqaman et al \cite{jaqaman_robust_2008} and Chenouard et al \cite{chenouard_multiple_2013}. Jaqaman et al write that the most accurate solution to single-particle tracking is provided by the method of multi-hypothesis tracking (MHT). In MHT, given particle positions in every frame, all particle paths within the bounds of expected particle behavior are constructed throughout the whole movie. The largest non-conflicting ensemble of paths is then chosen as the solution. Jaqaman et al attempt to approximate MHT by solving a two-stage linear assignment problem. Even so, they describe this method as computationally prohibitive even with tens of particles tracked over tens of frames. 

Fortunately, modern-day computers are sufficiently powerful to track millions of independent universes, each containing dozens or hundreds of cells. Parallelization further reduces the computational burden. Our method is trivially parallelizable---each universe evolves separately from all the others and can be placed in a separate thread. Thus, we are able to track up hundreds of cells over as many frames with little difficulty.


\section{Methods}

\subsection{Overview of our method}

Essentially we maintain a computer simulation of what is happening in the real world, and maintain synchronization between the two using feedback from the video. The simulation is programmed with rules meant to mimic only things that can happen in the real world. This disallows impossible events like bacteria vanishing into thin air, appearing out of nowhere, or merging---all of these are major and frequent problems with existing techniques.

In the current proof-of-concept implementation, the world is very simple: bacteria are confined to two dimensions; each bacterium $b$ is modeled as a rectangle of length $L_b$ with semi-circles at each end; $b$ has a position $(x_b,y_b)$, an orientation $\theta_b$, and is allowed one ``bend'' point of angle $\phi_b$ if it is long enough (longer than a global constant $B$). Each bacterium can split into two daughter cells at any time with a probability $p$ that is a function of its length; $p$ is zero if $b$ is too short or has split too recently. Finally, we allow bacteria to overlap slightly but there is a repelling ``force'' that pushes overlapping bacteria away from each other.
Clearly, in future work our model can be generalized to allow arbitrary shapes, a third dimension, etc.

All variables are then moved forward in time stochastically. Each bacterium in the simulation is allowed to move a little bit, rotate a little bit, grow in length a little bit, and possibly split into two daughter cells. Each of these changes are drawn from some probability density function. Currently we simply choose most of these ``delta'' changes uniformly at random from a small interval. This is not necessarily computationally efficient because many deltas are highly unlikely; a better method would be to draw our deltas from distributions that better match reality; luckily, as we can see in Figure \ref{fig:distributions} below, our simulation {\em provides} these distributions as output once we fit the motion of the simulations to the motion observed in the video. Future work will allow us to learn these distributions automatically, so as to automatically make the simulation more efficient as it progresses.

Since all the above variables evolve with time stochastically in our simulation, it is unlikely that any one realization of the simulated universe will track the real universe. To solve this problem, we {\em duplicate} the simulated universe many times, and then evolve each one forward independently towards the next video frame. The group of universes is called an {\it ensemble}. At the next video frame, we pick a small set of $K$ universes from the ensemble that best match the video, discard all the others, and then continue forward from those $K$ starting points to the next frame. Currently the simulation must be initialized manually with each bacterium's initial values chosen by eye; we expect that it will be fairly easy to automate this process as well.


\subsection{Video input}

\begin{figure}[htbp]
\centering
\captionsetup{justification=centering}
\includegraphics[width=\linewidth]{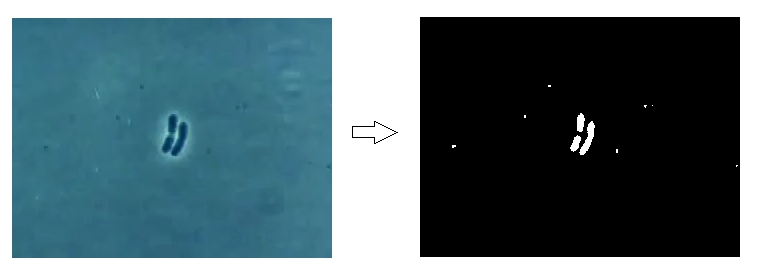}
\caption{We convert to binary image using simple brightness threshold.}
\label{fig:bin_image}
\vspace{2em}
\subcaptionbox{\centering ``Noisy" Frame 26}
{\fbox{\includegraphics[width=6cm]{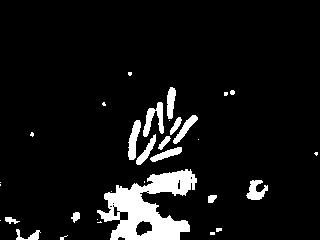}}} \hspace{1em}%
\subcaptionbox{\centering ``Clean" Frame 26}{\fbox{\includegraphics[width=6cm]{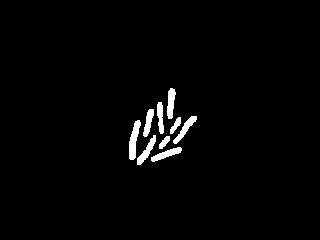}}}
\caption{(a) Simple thresholding can result in significant noise as seen in Frame 26. Although our algorithm is very robust even to such levels of noise, we also tested a more ``clean'' thresholding as seen in subfigure (b).}
\label{fig:noisy-clean}
\end{figure}

In order to set up our simulation, we first take as input a video of bacteria swarming and split it into its constituent frames\footnote{We chose a random video on YouTube, {\tt https://www.youtube.com/watch?v=UccyM8QeIeE}}. The cells in each frame are segmented using a simple intensity threshold. The resulting frames are binarized such that bacteria are represented in white and the background in black. Segmentation was tried both in Matlab and in Python using the opencv library. We deliberately do most of our tests on noisy binarized images without any additional clean-up or pre-processing (Figure \ref{fig:bin_image}); our method is sufficiently noise-resistant that additional pre-processing would be unnecessary and computationally wasteful, although we do use clean images (Figure \ref{fig:noisy-clean}) for comparison later.

\subsection{Simulation} 
\begin{figure}
\begin{lstlisting}[escapechar = \*]
Initialize: space S = {U}
for t = 0...end:
    new_S = * $\emptyset$ *
    for each universe U in S:
        add neighbors(U) to new_S
        evolve U to next frame
    pull all universes in new_S toward reality
    S = {top k universes closest to reality}

\end{lstlisting}
\caption{Pseudocode for our per-frame method}
\label{fig:pseudo}
\end{figure}

The pseudocode in Figure \ref{fig:pseudo} broadly describes each step of our simulation.
We begin the simulation by initializing the space $S_0$ to be a set that contains only one universe. This universe is a set of bacteria whose attributes, including coordinates, length, and orientation, are set manually. Then, we input the binarized frames from the video of real bacteria. In our experiment, we have 120 frames labelling ``time'' t running from 1 to 120. At each time $t$, we use the universe(s) in $S_t$ to generate possible futures and then compare the $(t+1)\textsuperscript{st}$ frame to each new universe, keeping only the universe(s) that are closest to the observed reality. These universes will be the new space $S_{t+1}$ for the next iteration at time t + 1.


Each simulation also keeps track of whether any two bacteria in its universe are overlapping, by using simple geometry to detect if the rectangles or semi-circle ends of each bacterium overlaps its neighbors. Currently this is done with a simple loop over {\em all} bacteria, which is computationally inefficient; methods such as k-d trees \cite{samet1990design} could readily be implemented to reduce this collision detection cost.

\subsection{Comparing simulated universes to the real one}
To compare the real to our simulated universes, each simulation outputs a synthetic image of the state of its bacterial colony. We compare each synthetic image to the real binarized image and perform a simple image subtraction; the size of the residual after image subtraction is our cost function. An exact match would be a blank residual image.
The objective of the simulation is to minimize the cost function at each frame; a more global sum of cost functions is another possibility we explore below.

To compute the cost, we load the binarized images into the software as matrices. This allows us to use efficient vectorized operations. We then compute the absolute elementwise difference between the real and simulated matrices for each frame. The frames with the fewest differences have the lowest cost. It is, of course, worth noting that the universe with the lowest cost according to the cost function may nevertheless not appear to fit the bacteria in the real frame as well as a universe with a slightly higher cost. Because the cost function cannot {\em perfectly} measure the difference between the simulation and reality, we retain $K$ of the best-match universes. It may also be advantageous to enforce a diversity criterion onto which universes we keep. For example, if all $K$ best batch universes represent the exact same family tree and only differ in minor position and rotation of each member, that may be less desirable than forcing us to keep some slightly-less good image matches in favor of diversifying the possibilities in terms of cell family trees. This is an area of future work.



\subsection{Pulling stage}

\begin{figure}[htp]
\centering
\captionsetup{justification=centering}
\includegraphics[width=\linewidth]{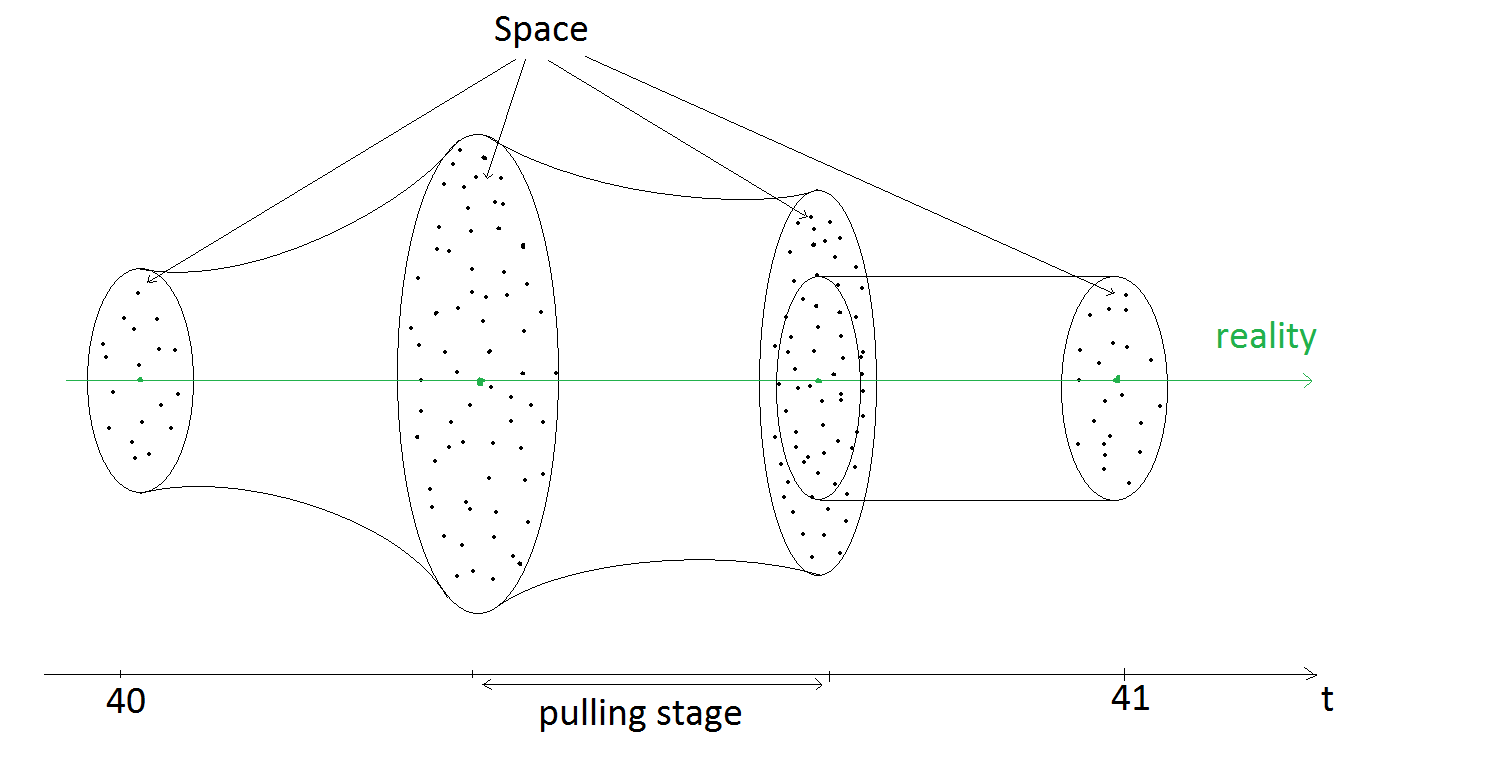}
\caption{Pulling Stage}
\label{fig:Pulling_Stage}
\end{figure}

After generating many possible futures and retaining the best matches among them, we ``pull" those best matches even closer to the real universe (see Figure \ref{fig:Pulling_Stage}). In the pulling stage, we attempt to bring each best-match universe as close to reality as possible. We accomplish this by iterating over each universe and checking whether we can lower its cost by moving or growing (or possibly shrinking) each individual bacterium in it one unit at a time. If the change does in fact lower the cost, we apply it to the universe. We continue to incrementally change each bacterium in the universe until we can make no changes that lower the universe's cost. Thus, at the end of this stage, all the best-match universes will have reached their respective local minimum costs. This ensures that the universes comprising our new space $S_{t+1}$ are of the highest possible quality.

\subsection{Time and Space Complexity} 
The program takes the most time when generating new universes and their costs, so running time complexity is based on the total time the simulation creates universes.

\begin{lstlisting}[escapechar = \*]
Let the number of bacteria be N.
Cost function: O(N)
For each frame at time t:
    generate about F future universes per bacterium, 
    N bacteria per universe, and 
    there are K universes in the space S.
=> Total time for generating universes: FKN
=> Total running time: O(FKN*$^2$*)
\end{lstlisting}

\subsection{Traceback stage}
\label{sec:traceback}

Because of the design of our simulation, the best-fitting universe at step $t+1$ may not have come from the best-fitting universe at time $t$. This means that the sequence of best-fitting universes may not form a consistent set representing one possible universe. In particular, we have observed that, while following a sequence of best-fitting universes from frame-to-frame, it can sometimes occur that two bacteria at time $t$ reunite into one bacterium in frame $t+1$; this is, of course, impossible in the real world (and therefore never happens in any one of our simulated universes), and visually happens only because the best frame at $t+1$ did not come from the best-fitting one at time $t$. Accordingly, to construct consistent lineage trees, we first need to construct a single {\em universe} that is consistent from the beginning to the end of the video. Luckily, every universe at the last frame came from {\em some} universe in the previous frame, so we are able to pick any one universe at the end and trace it backwards through time back to frame 0. This provides one self-consistent universe from beginning to end.

\section{Results}

Our objective is to accurately maintain a correct picture of reality as depicted in the video scene. Existing methods are often able to statistically track a swarm of cells, which provides measures such as an approximate count and gross movement properties. However there are measurements that can be made with detailed individual tracking that would be impossible with a less rigorous approach.

\begin{figure}[t]
\centering
\captionsetup{justification=centering}
\includegraphics[width=\linewidth]{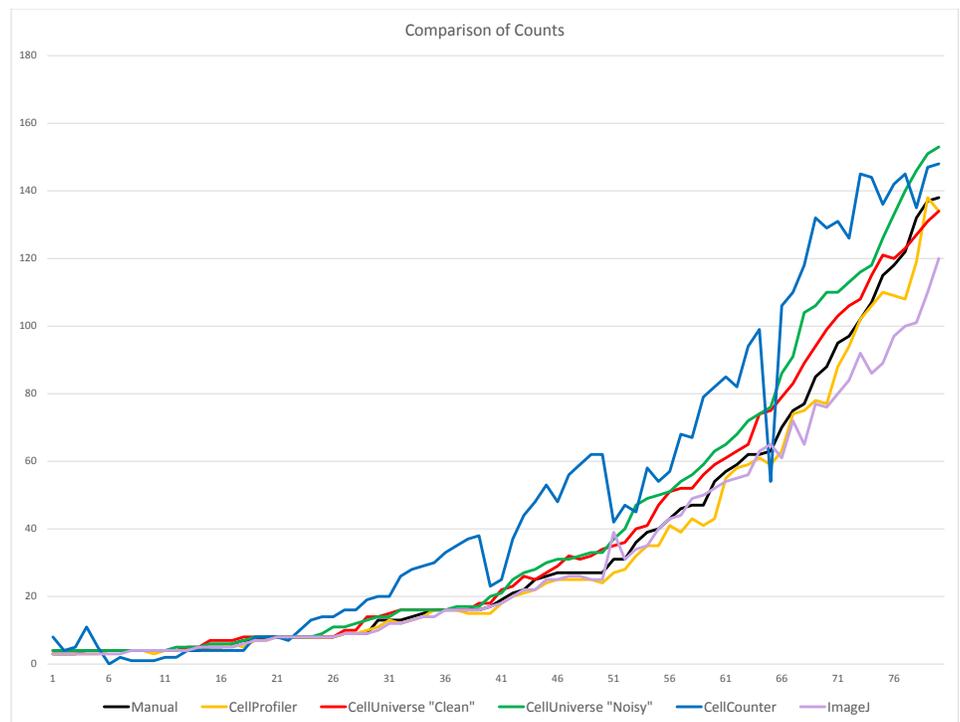}
\caption{The comparison of the counts returned by each of the methods.}
\label{fig:counts}
\end{figure}

Below we demonstrate how our conglomerate counts of bacteria as a function of time (Figure \ref{fig:counts}) have an ``error'' compared to human counts that is comparable with existing methods (Figure \ref{fig:count-error} and Table \ref{tab:counts}). Then, we demonstrate some measures that would be impossible without the perfect individual tracking that we offer. Such measures include a family tree of every visible bacterium, as well as detailed statistics on motion, shape change, and a precision in reproduction rates that allows us to unequivocally show that the cells at the periphery are reproducing at rate slightly but measurably faster than those near the center---a measurement that would be impossible without a precise family tree that demonstrates that the cells at the periphery are one generation ahead of the rest of the colony.

\subsection{Bacterial Counting}

\begin{figure}
\centering
\captionsetup{justification=centering}
\includegraphics[width=\linewidth]{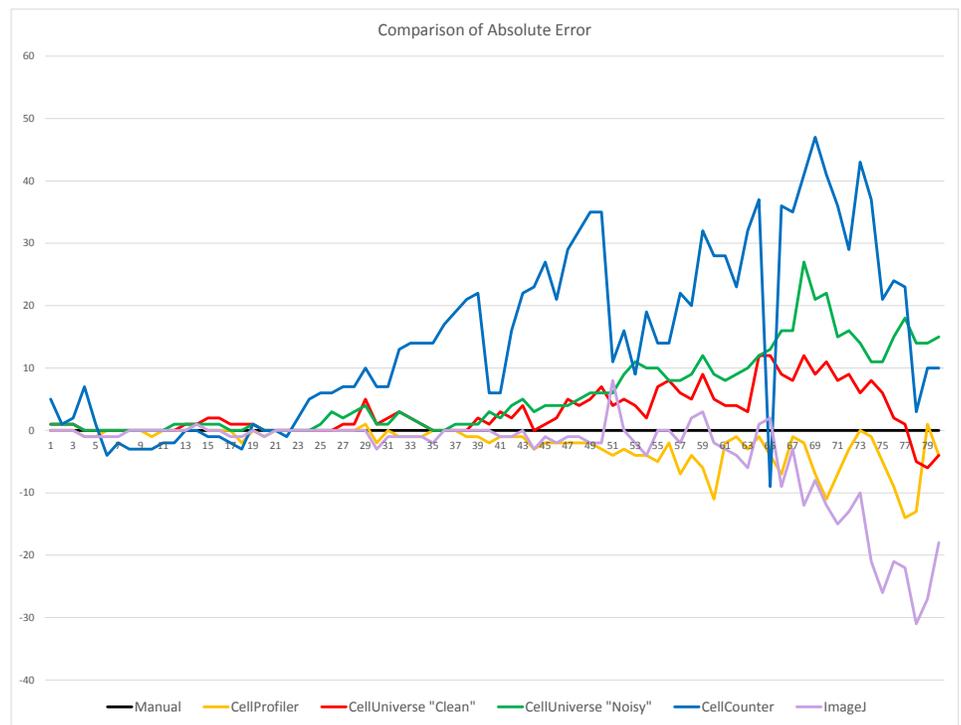}
\caption{The absolute error made by each method on the selected video. Our method tended to overestimate the number of cells in later frames (see {\em Limitations and future work}).}
\label{fig:count-error}
\end{figure}

To evaluate the quality of our simulation’s counts, we ran our simulation on two sets of binarized frames: a “noisy” set and a “clean” set. We used both sets in order to demonstrate that our software is not impeded even by high amounts of noise as depicted in Figure \ref{fig:noisy-clean}. Then, we compared the frame-by-frame bacterial counts produced by our simulation to frame-by-frame bacterial counts we conducted manually on the original video. For instructive purposes, we also compare our software’s performance to the performance of a few other major cell counting programs---CellCounter, ImageJ, and most significantly, CellProfiler---on the same sample video. Results are presented in Table \ref{tab:counts} and depicted in Figures \ref{fig:counts}--\ref{fig:count-error}.

\begin{table}
\begin{center}
\begin{tabular}{ |p{5cm}||p{2.5cm}|}
 \hline
 \multicolumn{2}{|c|}{Average Error of Counts} \\
 \hline
 Program& Average Error\\
 \hline
 CellProfiler & -2.20\\
 Cell Universe ``Clean"&  2.85 \\
 ImageJ & 3.98\\
 Cell Universe ``Noisy"   & 5.85\\
 CellCounter    & 14.56\\
 \hline
\end{tabular}
\end{center}

\begin{center}
\begin{tabular}{ |p{5cm}||p{2.5cm}|}
 \hline
 \multicolumn{2}{|c|}{Overall Percent Error} \\
 \hline
 Program& Percent Error\\
 \hline
 CellProfiler & -6.17\\
 Cell Universe ``Clean"&  7.99 \\
 ImageJ & -9.95\\
 Cell Universe ``Noisy"   & 16.40\\
 CellCounter    & 40.82\\
 \hline
\end{tabular}
\end{center}
\caption{Absolute and relative count errors compared to manually counting.}
\label{tab:counts}
\end{table}

\begin{figure}[htp]
\centering
\subcaptionbox{\centering Video Frame 67}
{\fbox{\includegraphics[width=6cm]{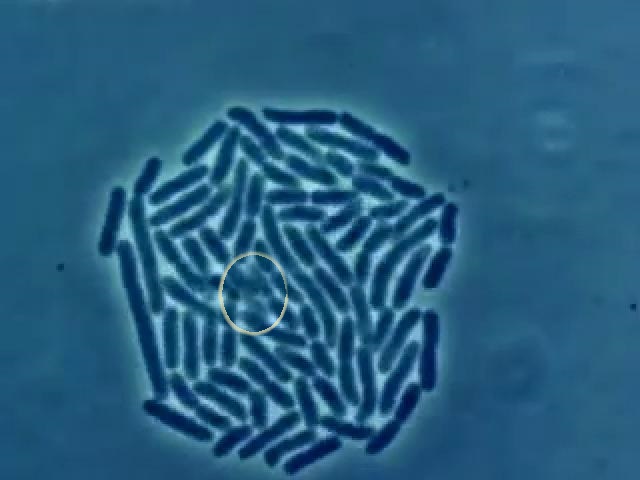}}} \hspace{1em}%
\subcaptionbox{\centering Video Frame 68}{\fbox{\includegraphics[width=6cm]{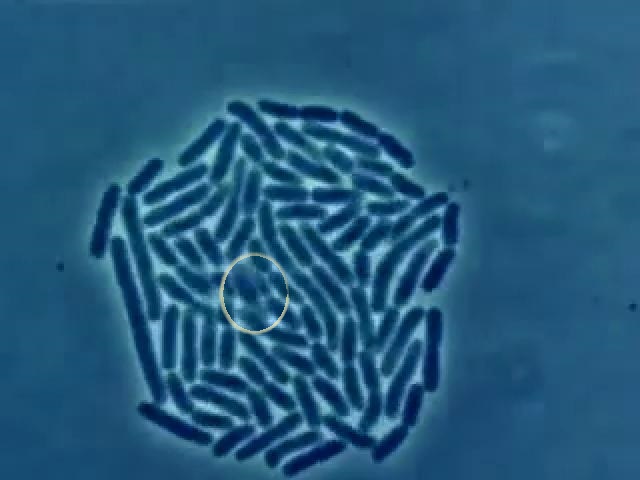}}}
\subcaptionbox{\centering ``Noisy" Frame 67}
{\fbox{\includegraphics[width=6cm]{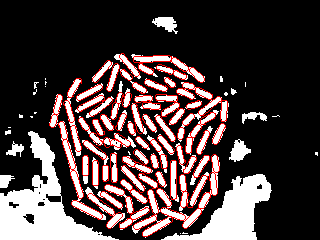}}} \hspace{1em}%
\subcaptionbox{\centering ``Noisy" Frame 68}{\fbox{\includegraphics[width=6cm]{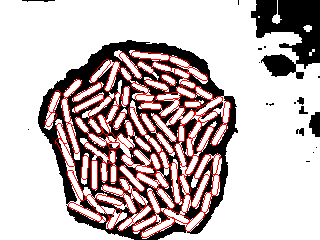}}}
\subcaptionbox{\centering ``Clean" Frame 67}
{\fbox{\includegraphics[width=6cm]{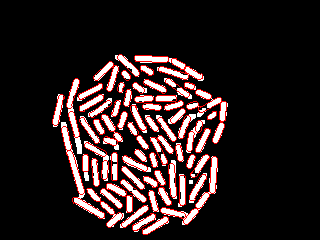}}} \hspace{1em}%
\subcaptionbox{\centering ``Clean" Frame 68}{\fbox{\includegraphics[width=6cm]{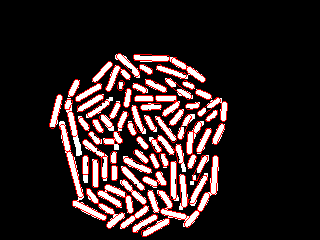}}}


\caption{Between Frames 67 and 68, our software begins to encounter difficulty tracking cell movements in the center of the colony, which is understandable since even a human may have difficulty discerning what has happened in the circled region of the video frames (subfigures (a)--(b)). Our lenient, ``noisy'' thresholding (subfigures (c)--(d)) arguably does a better job of segmentation in that region than the more aggressive ``clean" thresholding ((e)--(f)), which completely loses the blurry cells in the middle.}
\label{fig:1st-fuckup}
\end{figure}

Our simulation was able to {\em exactly} count the number of cells in each frame at least until Frame 66. Starting in Frame 67 (see Figure \ref{fig:1st-fuckup}), problems with the real video began to impact the quality of our simulation's count (and the counts of the other programs). The real video became blurry, and crowding at the center of the colony began to push certain individual bacteria off the focal plane under the mass of the colony. As a result, neither our simple ``clean" nor ``noisy" binarizations were entirely able to capture all of the bacteria actually “in” the frame. The combination of blurring and crowding made it difficult for the cost function to measure the difference between the simulation and reality.

In spite of these issues, our software achieved counting results superior to ImageJ’s and CellCounter’s, and comparable with CellProfiler’s. It bears mentioning that CellProfiler produced these results only after some experimentation with the plenitude of parameters available in the IdentifyPrimaryObjects module. While it is possible that the difficulties we experienced with CellProfiler might have been trivially surmounted by an expert, we emphasize that our own simulation requires only a single, extremely simple binarization, and that even extremely noisy binarizations did not unduly affect the quality of our results. 

\begin{figure}[htb]
\centering
\captionsetup{justification=centering}
\includegraphics[width=.8\linewidth]{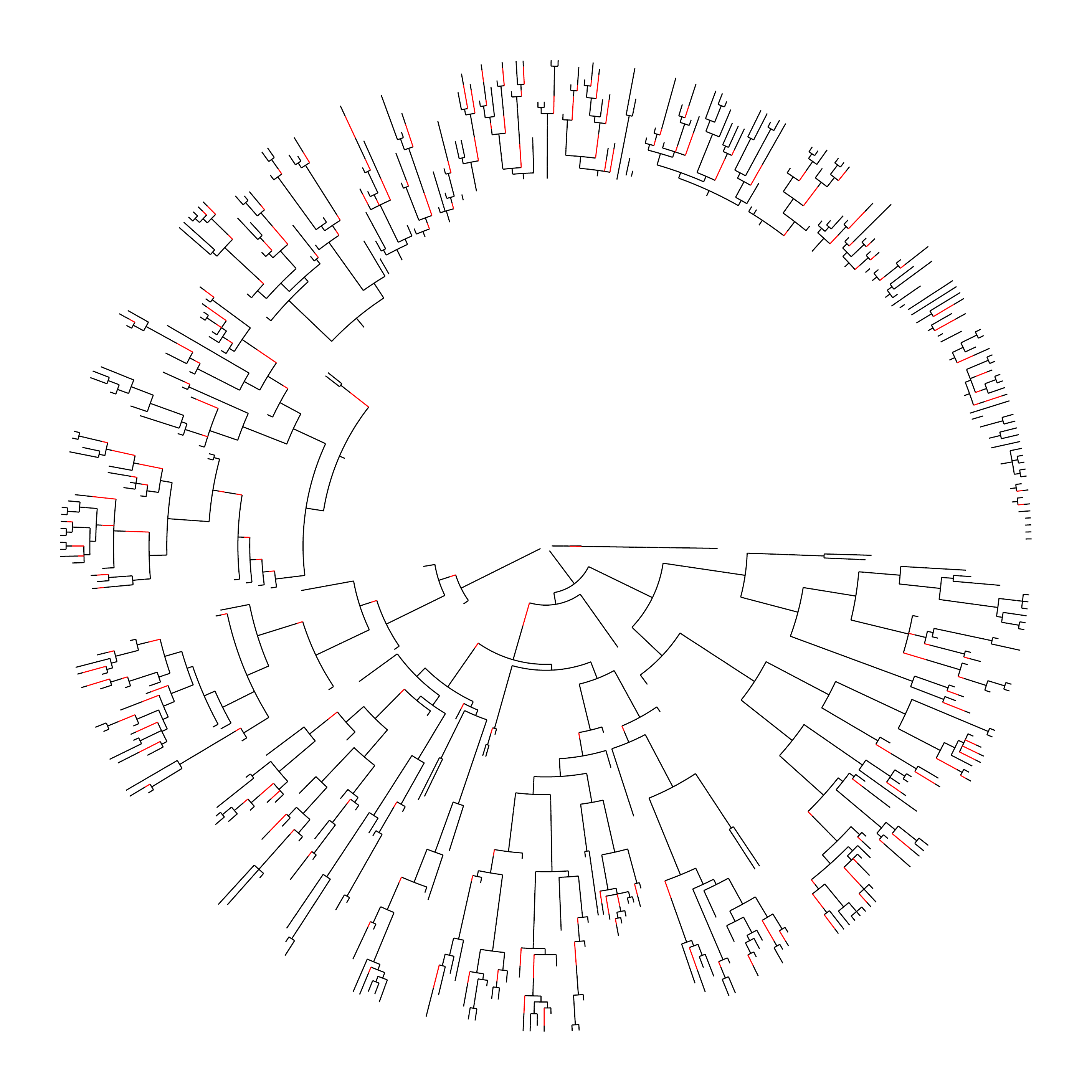}
\caption{CellProfiler LAP Tree. Time increases outward from the center. We note many inconsistencies. First, all cells must descend from one of the cells in the center of the image (those that existed at the beginning of the video), not appear out of nowhere as many, many cells do here; second, every leaf of the family tree should reach the edge of disk, which represents the end of the video, rather than vanish out of existence before reaching the outer edge as they do here.}
\label{fig:lineageCP-LAP}
\end{figure}

SUPERSEGGER WORK BEGINS HERE
{\em NOTE: I haven’t extracted official count results yet. Stand by for details. I will integrate this section fully when we receive comments. }

For thoroughness, we also compared our method to the most promising newly-published cell-tracking software we discovered in our review of the literature: SuperSegger, created by Stylianidou \cite{stylianidou_supersegger:_2016}. Unlike CellProfiler and ImageJ, SuperSegger is an automated MATLAB-based image processing package, not standalone software. It segments cells using a modified watershed algorithm assisted by neural networks to determine boundaries. Like our software, SuperSegger is optimized for rod-like bacterial cells and can both count and track cell lineages over many generations. Moreover, SuperSegger seems to be the most accurate of similar recently-published cell-tracking software packages, such as Oufti \cite{paintdakhi_oufti:_2016}. Accordingly, we believed a comparison of our method to SuperSegger would be instructive. 
	
	Counts
(description of counts to come) 

\begin{figure}[htb]
\centering
\captionsetup{justification=centering}
\includegraphics[width=\linewidth]{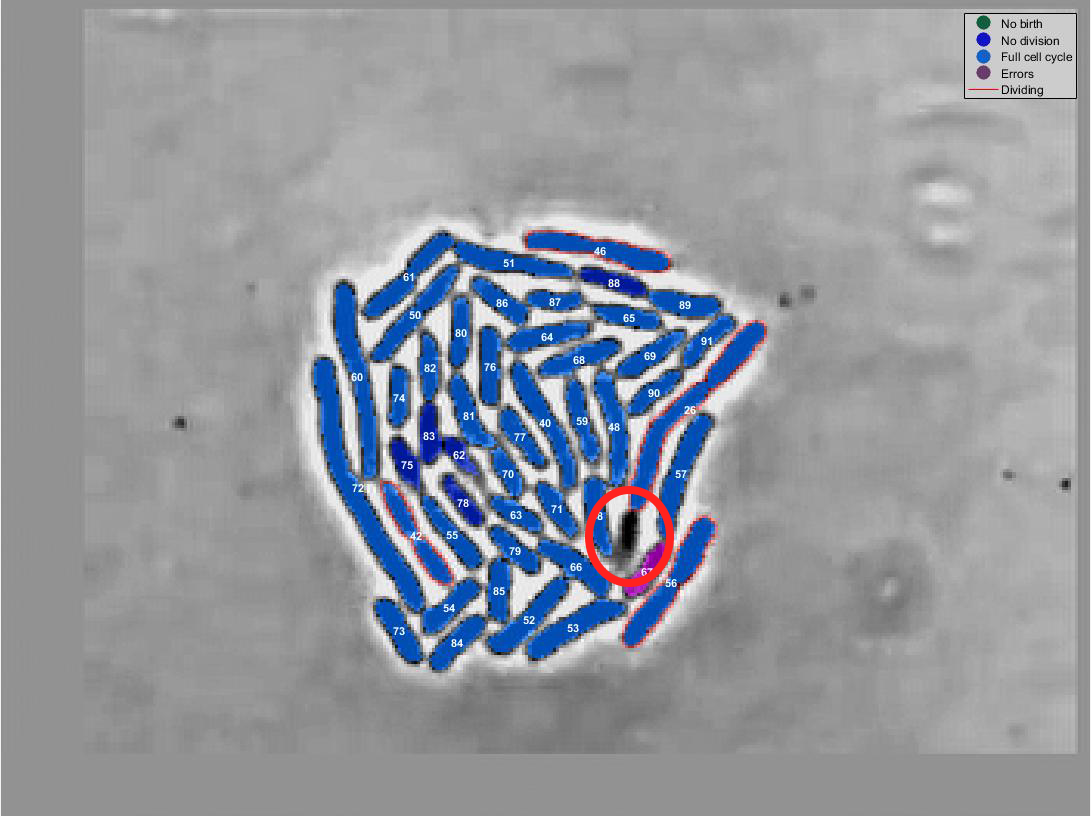}
\caption{This is frame 58 from the SuperSegger output. Although the cell inside the red circle is properly recognized as a cell in frames 57 and 59, here it is not.}
\label{fig:ss58}
\end{figure}

\begin{figure}[htb]
\centering
\captionsetup{justification=centering}
\includegraphics[width=\linewidth]{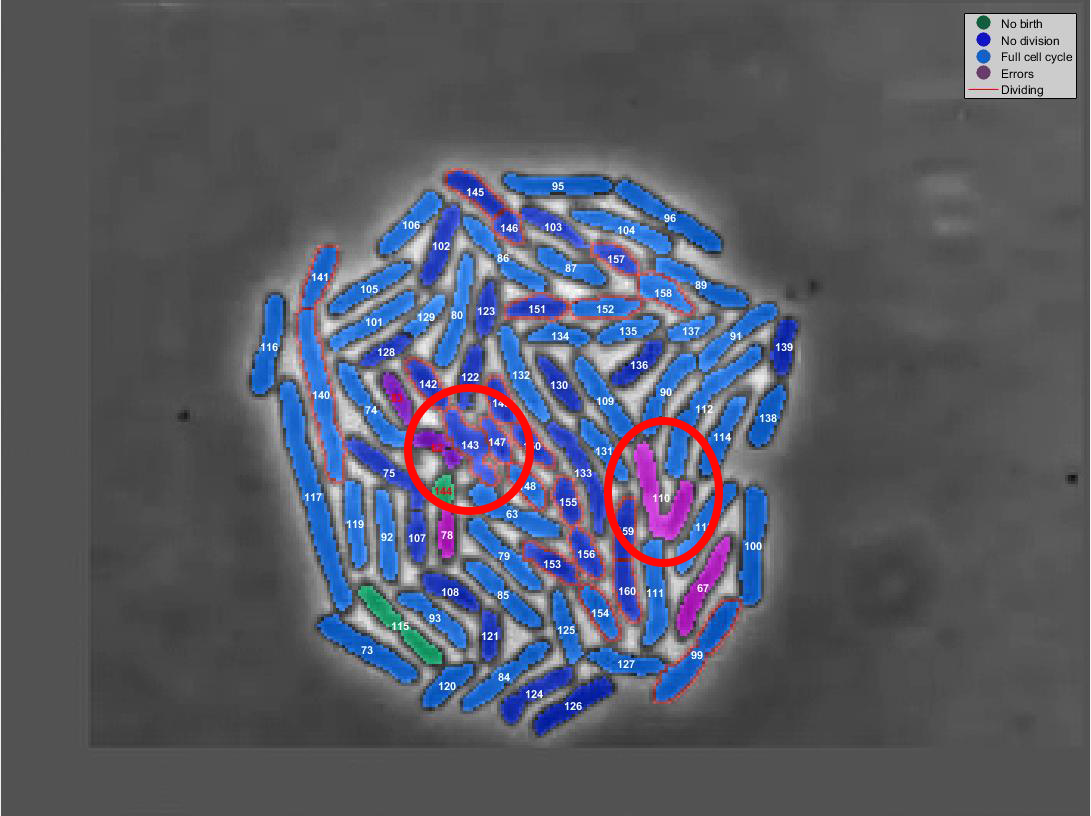}
\caption{This is frame 67 from the SuperSegger output. The circled areas contain cells that are obviously not rods. The cell on the left is a jagged agglomeration of two or three different cells; the cell on the right is simply two cells whose southern ends overlap somewhat. It is surprising to see such errors in software optimized for rod-shaped bacteria. Physical modeling manages to trivially avoid errors like these.}
\label{fig:ss67}
\end{figure}

\begin{figure}[htp]
\centering
\subcaptionbox{\centering SuperSegger Frame 75}
{\fbox{\includegraphics[width=6cm]{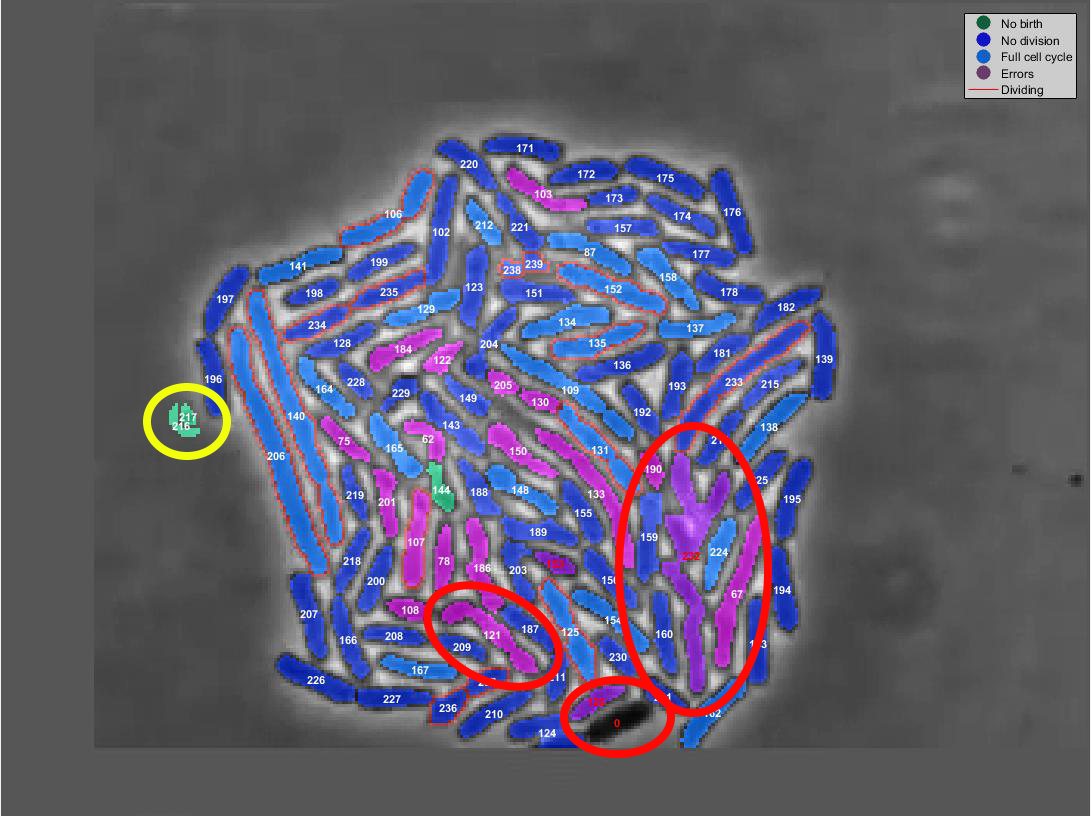}}} \hspace{1em}%
\subcaptionbox{\centering SuperSegger Frame 76}
{\fbox{\includegraphics[width=6cm]{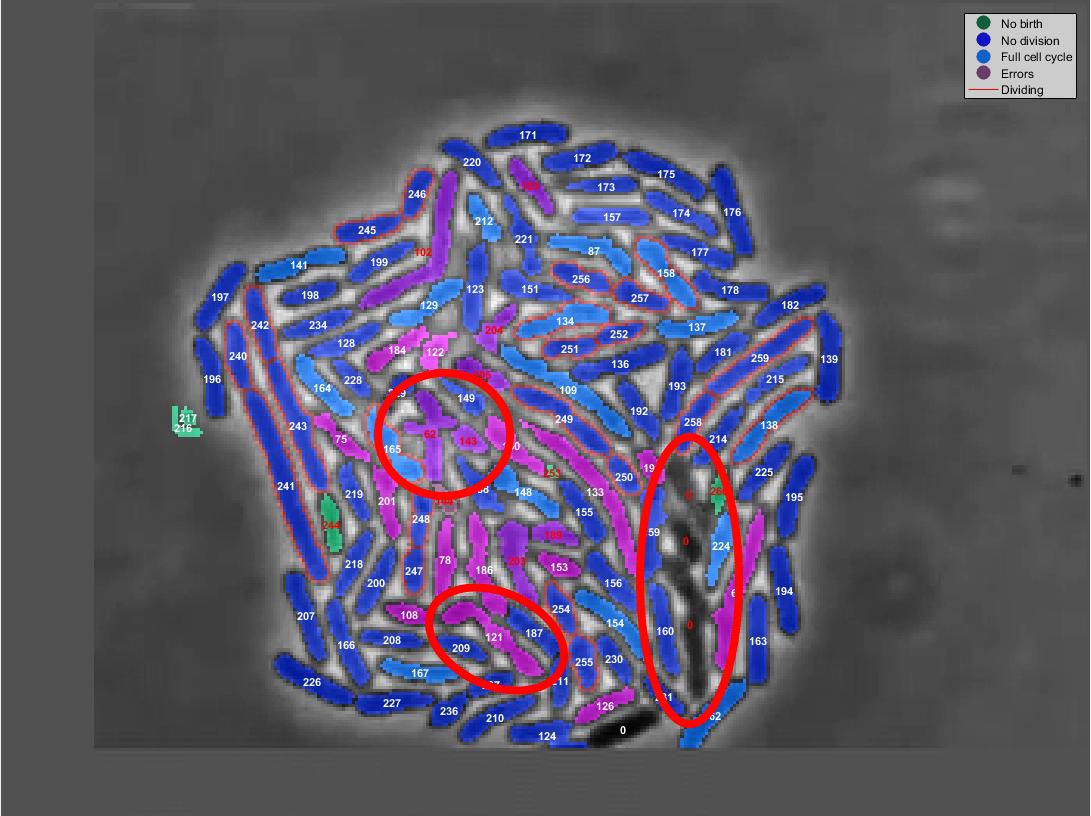}}}
\caption{In Frame 75, the areas circled in red are additional oddly-shaped merges and uncounted cells. The area in yellow is noise; SuperSegger interprets this noise as a new cell, and updates its lineage tree accordingly--and mistakenly. In frame 76, the areas circled in red on the left are oddly-shaped merges. On the right, it is possible to see several cells that SuperSegger simply stopped counting.}
\label{fig:ss70errors}
\end{figure}

Manual review of the output frames from SuperSegger reveals it manages to track cells without error up until frame 55. At that point, several odd counting errors become evident. For example, a cell vanishes from SuperSegger’s count between frames 57 and 59 despite the cell still obviously being there (see Fig. \ref{fig:ss58}). At frame 67, SuperSegger begins to merge several individual cells into strange agglomerations, counting the merges as cells in their own right despite the fact that they are plainly not rod-shaped (see Fig. \ref{fig:ss67}). These merged-cell errors accumulate until frame 75, at which point several cells are dropped altogether (see Fig. \ref{fig:ss70errors}). 

Lineage Tracking

\begin{figure}[htb]
\centering
\captionsetup{justification=centering}
\includegraphics[width=\linewidth]{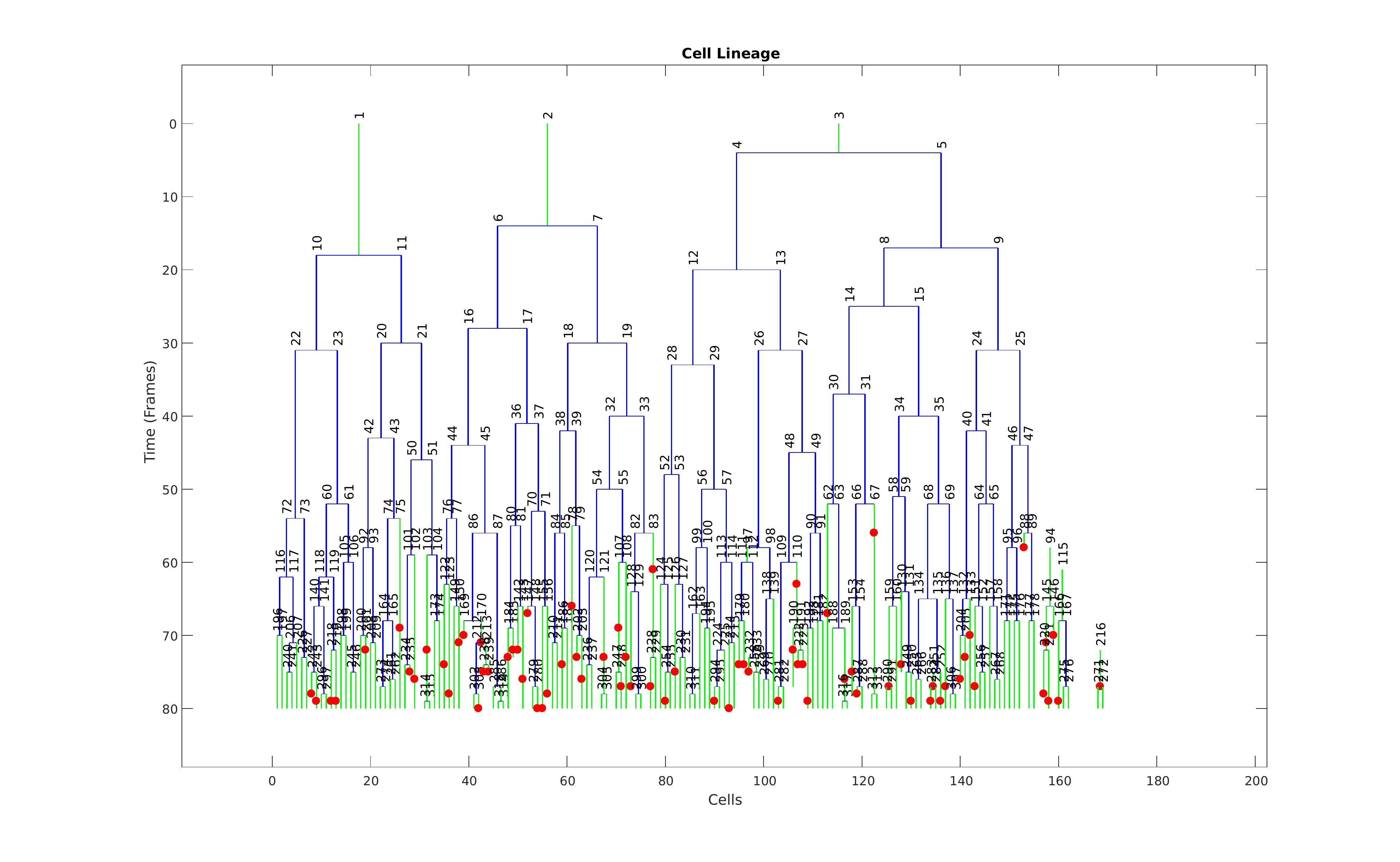} 
\caption{SuperSegger Lineage Tree. Note the branches on the far right, which appear spontaneously several generations into the video.}
\label{fig:SSLineageTree}
\end{figure}

Fig. \ref{fig:SSLineageTree} demonstrates SuperSegger’s fairly-robust built-in lineage tracking functionality. Unlike CellProfiler’s distance method, SuperSegger manages to track cells fairly accurately from frame-to-frame; no cells instantly split again after they have already split, and all splits are into two cells rather than three or more. However, SuperSegger makes two errors similar to those made by CellProfiler’s LAP method, “discovering” three cells that appear from seemingly nowhere fairly late into the colony's lifetime. 

SUPERSEGGER WORK ENDS HERE

\subsection{Lineage Tracking}

Our simulation is also capable of tracking the lineage of every single bacterium it observes. In the simulation, we give each bacterium a binary name, and whenever it splits and produces two daughters, we name them by adding a zero and a one to its parent's name. For example, a bacterium named '00' will split into two children named '000' and '001'. We store the name, position, and orientation of every bacterium in a universe in a corresponding text file. This naming system enables us to construct a lineage tree of all bacteria at the end of the simulation. Note that for this to work we first build a self-consistent universe as described in \S \ref{sec:traceback}.

\begin{figure}[htb]
\centering
\captionsetup{justification=centering}
\includegraphics[width=0.8\linewidth]{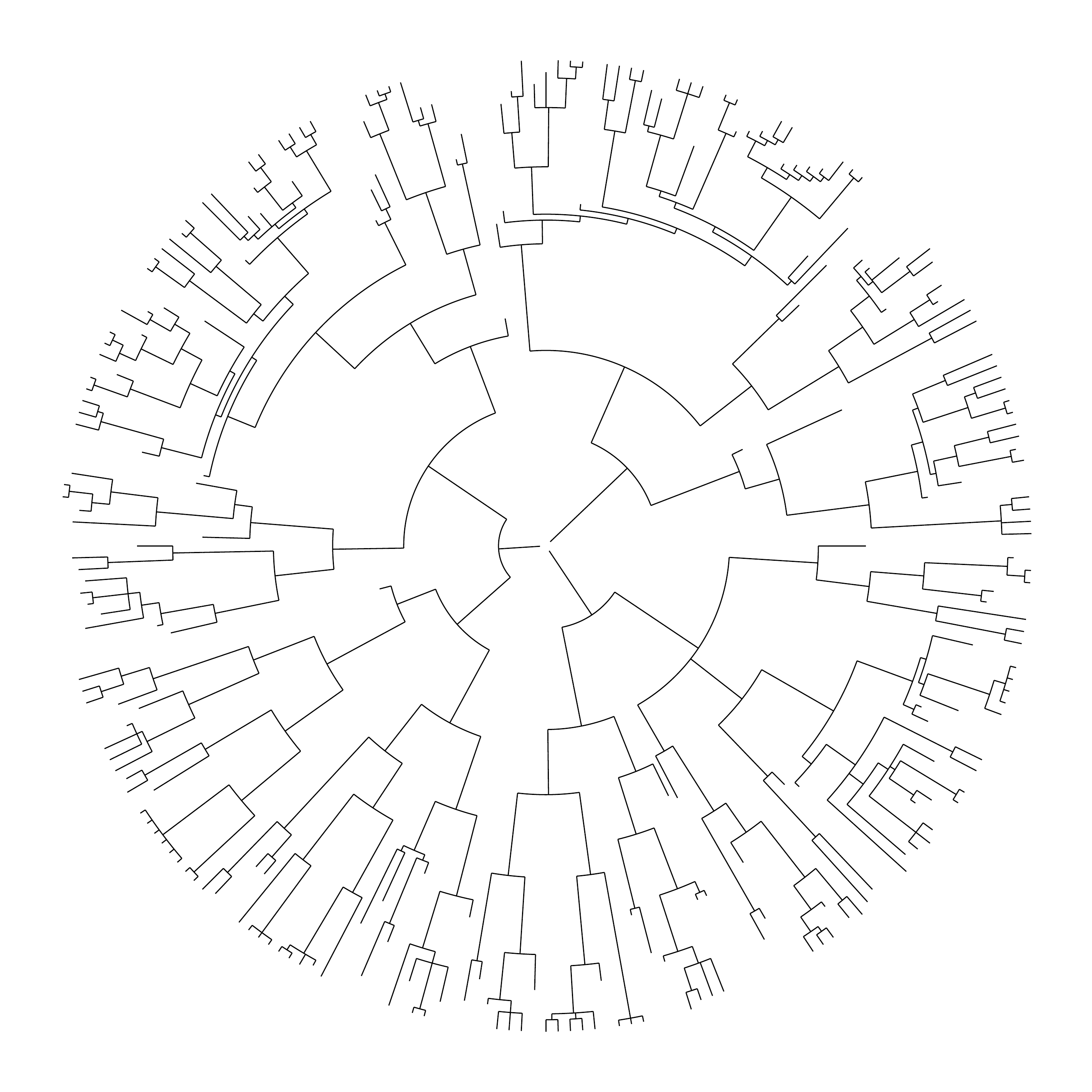}
\caption{CellProfiler Distance Tree. Although better than the LAP tree, we note different inconsistencies here. First, in the real world one bacterium can split into two, but it cannot split into 3 or more as can be seen in several places; second, each leaf from the family tree should reach the edge of the disk rather than stop somewhere inside the disk.}
\label{fig:lineageCP-Dist}
\end{figure}

\begin{figure}[htp]
\centering
\subcaptionbox{\centering Frame 48 (counting output)}{\fbox{\includegraphics[width=6cm]{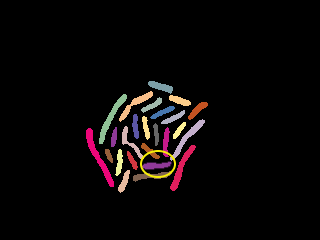}}} \hspace{1em}%
\subcaptionbox{\centering Frame 49 (counting output)}{\fbox{\includegraphics[width=6cm]{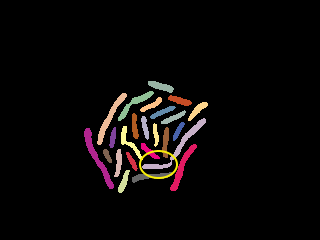}}}
\subcaptionbox{\centering Frame 48 (tracking output)}{\fbox{\includegraphics[width=6cm]{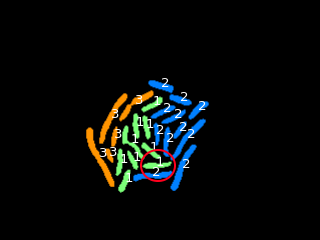}}} \hspace{1em}%
\subcaptionbox{\centering Frame 49 (tracking output)}{\fbox{\includegraphics[width=6cm]{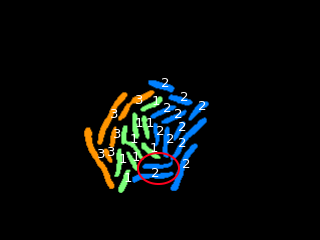}}}
\caption{These output images from CellProfiler illustrate the difficulty it encounters with single-pixel bridges. A single-pixel bridge between the cell outlined in yellow/red and one of its neighbors was treated as a merge. When the cells moved apart, CellProfiler counted a parent-daughter split.}
\label{fig:CP-bridges}
\end{figure}

\begin{figure}[htb]
\centering
\captionsetup{justification=centering}
\includegraphics[width=0.8\linewidth]{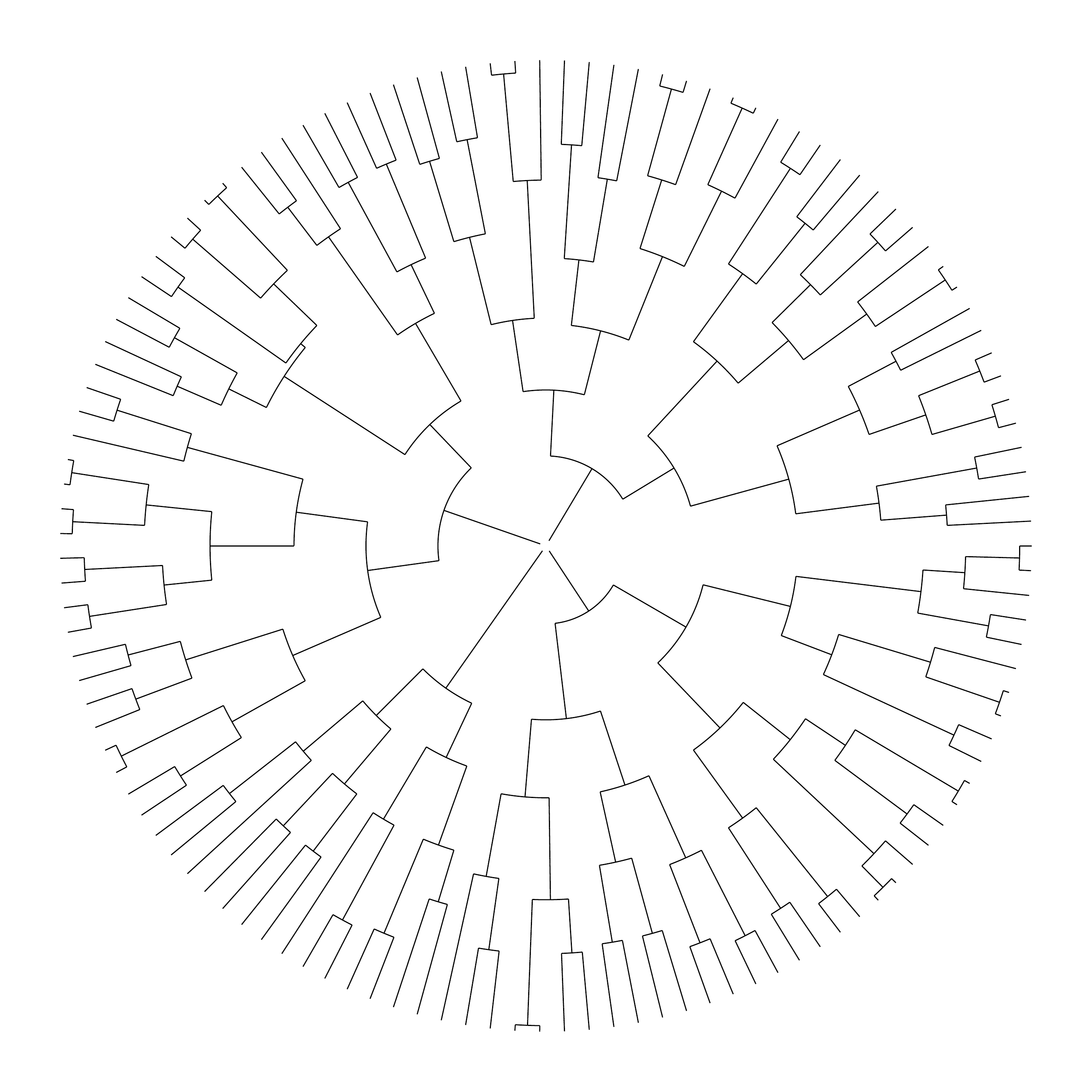}
\caption{Cell Universe Lineage Tree. Every bacterium is a descendent of one of the initial 4 starting at the center (nobody appears out of thin air); cells can only split into 2, never 3 or more; and every branch of the tree reaches the perimeter of the disk (because nobody vanishes out of existence). Using this depiction one can get the following observations: an accurate cell count at any time $t$ by drawing a circle of radius $t$ and counting the number of lines that intersect said circle; an estimate of the mean lifetime of bacteria in this colony by averaging the lengths of all the branches; and by counting the edges, we can observe that most bacteria at the edge of the disk are in generation 6, with just a few in generation 7.} 
\label{fig:lineageCU}
\end{figure}

Of the cell counting programs we examined, only CellProfiler is capable of tracking objects as they split and multiply. However, our simulation was consistently able to produce more complete and meaningful lineage trees for the bacteria. Figures \ref{fig:lineageCP-LAP}--\ref{fig:lineageCU} are radial lineage trees. The initial cells are represented by the branches at the very center of the tree; as time elapses, they split outward from the center into new branches. Note, however, that the tree does not depict physical locations on the plate, the layout is based instead on maintaining a roughly constant density of lines for a circle of any given radius centered on the center of the disk.

Figure \ref{fig:lineageCP-LAP} is the lineage tree generated based on the output of the Linear Assignment Problem (LAP) method in CellProfiler's TrackObjects module. The LAP method attempts to link objects between consecutive frames and solve a global combinatorial optimization problem to determine their most likely trajectories, and is based on research by Jaqaman et al \cite{jaqaman_robust_2008}. The red branches represent cells that were lost but eventually found again. Notwithstanding the numerous red lines (and the numerous lost cells they represent) the LAP method also seems to have been able to trace lineages for only around half the cells in the video. The other half appear from nowhere well into the video, some apparently in the very last frame. Moreover, according to the LAP method, one of the original cells in the very first frame of the video never split! 

CellProfiler's distance method, whose lineage tree appears in Figure \ref{fig:lineageCP-Dist}, tracks objects much better than the LAP method, but it is not without its flaws. At several places, the distance tracking method captures splits of cells into more than two daughters--usually three, but as many as six. Moreover, several cells apparently do not persist until the end of the simulation. As far as we were able to ascertain by manually curating every frame of the video, at no point do any of the bacteria split into more than two daughters---and at no point do cells vanish outright. We believe these tracking errors were precipitated by CellProfiler's inability to effectively distinguish between closely-packed objects. The cells in our video aggregate closely enough that they frequently touch edges and thus appear to be connected, sometimes by single pixels. As Figure \ref{fig:CP-bridges} illustrates, CellProfiler merges objects that are connected, even by single pixels. This seems to remove them from the tracking, resulting in several prematurely-terminated branches in the lineage tree. When these single-pixel bridges are severed by bacterial motion, the merged ``object"---which was really always two or more separate objects---appears to suddenly split into many daughters. The combination of these problems yields the irregular, inaccurate lineage tracking represented in Figure \ref{fig:lineageCP-LAP}. 

Our own lineage tree appears in Figure \ref{fig:lineageCU}. It reflects that our method does not lose any members, tracks every cell that existed at the beginning of the simulation {\em completely} until the end of the simulation, and captures binary splits---the only kind we observed in our manual curation. We think it worth reiterating at this point that by virtue of our reliance on physical rules, our model does not make any of the mistakes made by the CellProfiler methods. It is, after all, physically impossible (and blatantly contradicted by reality) that over half the cells in the video suddenly appeared from nowhere. Moreover, although it is not {\em impossible} that one of the cells in the video split into three daughters (though we did not observe this), we find it essentially impossible that any of the cells split into {\em as many as six} all at once. Recall that these splits into six cells were precipitated by spurious merges of cells whose edges touched. It is difficult to imagine any scenario where a one-pixel bridge between objects in general (let alone bacteria specifically) reflects an genuine merge of those objects in reality; the physical rules on which our simulation relies reflect that. We are thus able to accurately track bacteria even when they are packed edge-to-edge.

\subsection{Novel measurements only possible with our method}

\subsubsection{Reproduction rates as a fuction of position}
\begin{figure}[htp]
\centering
\captionsetup{justification=centering}
\includegraphics[width=12cm]{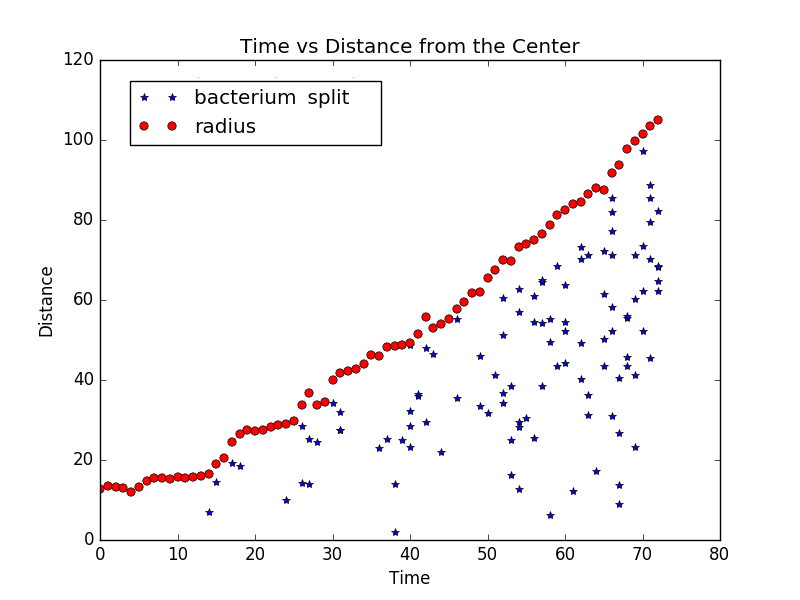}
\caption{Scatter plot showing how far each bacterium was from the center of the colony as a function of the time that it split. Red dots represent whichever bacterium is furthest from the center. Careful inspection of the far right of the figure shows that more splitting is occurring far from the center (above about distance 60) than closer to the center; the family tree shows that these bacteria are exactly one generation ahead of those at the center, demonstrating that the reproduction rate is slightly higher at the periphery (by about 14-17\% since frame 70 contains bacteria in the 6th and 7th generations), possibly due to resource depletion near the center of the colony.}
\label{fig:Scatter_Plot}
\end{figure}

By keeping coordinates, orientations, lengths, and precise timing of cell division events of all of bacteria from the beginning of the simulation, we can use the information to learn more about cell behavior both individually and in conglomerate. For example, we can make a plot of distance of a bacterium from the center of its colony as a function of time where it starts splitting (Figure \ref{fig:Scatter_Plot}).

As the far right of Figure \ref{fig:Scatter_Plot} illustrates, bacteria near the edge reproduce before those near the center. For example, the bacteria along the edge of the colony near frame 74 are one full generation ahead of the rest of the rest of the colony, suggesting reproduction occurs faster near the periphery (the effect is made more obvious by the video included in our supplemental materials). This effect would be virtually impossible to detect without reliable tracking of every individual in the colony and the resulting highly accurate family tree. We hypothesize that this effect is caused by resource depletion in the center of the colony, which slows the growth of the bacteria in the center relative to those along the less-crowded edges. 

\subsubsection{Population statistics of individual motion and reproductive rate}
\begin{figure}[htp]
\centering
\subcaptionbox{\centering Movement in x-direction (pixel)}{\fbox{\includegraphics[width=6cm]{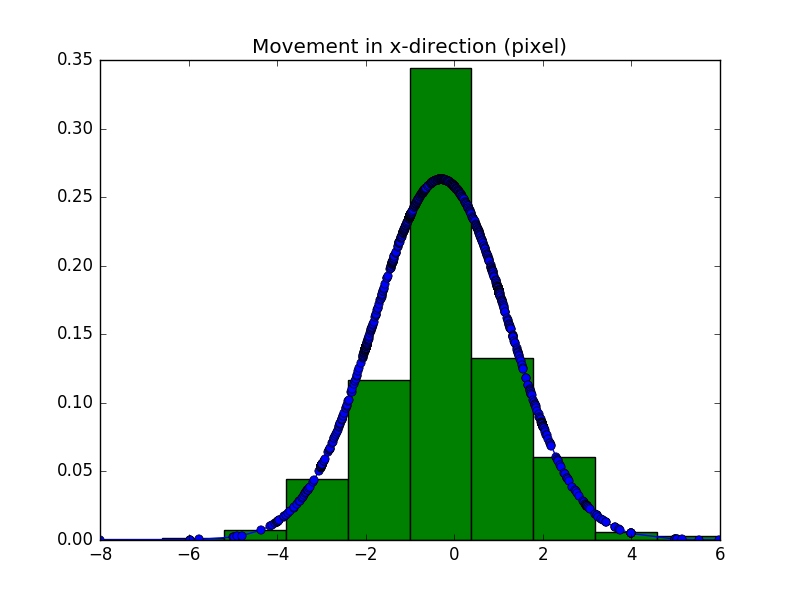}}} \hspace{1em}%
\subcaptionbox{\centering Movement in y-direction (pixel)}{\fbox{\includegraphics[width=6cm]{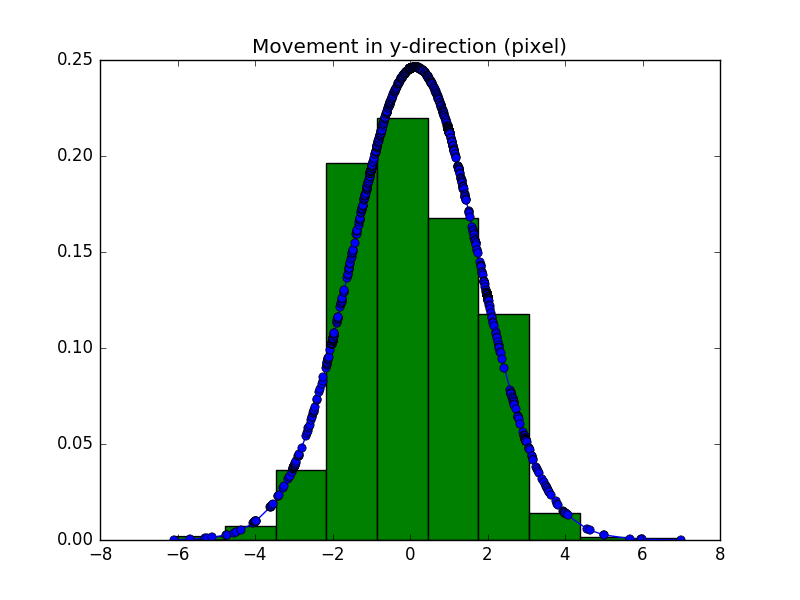}}}
\subcaptionbox{\centering Rotation (degree)}{\fbox{\includegraphics[width=6cm]{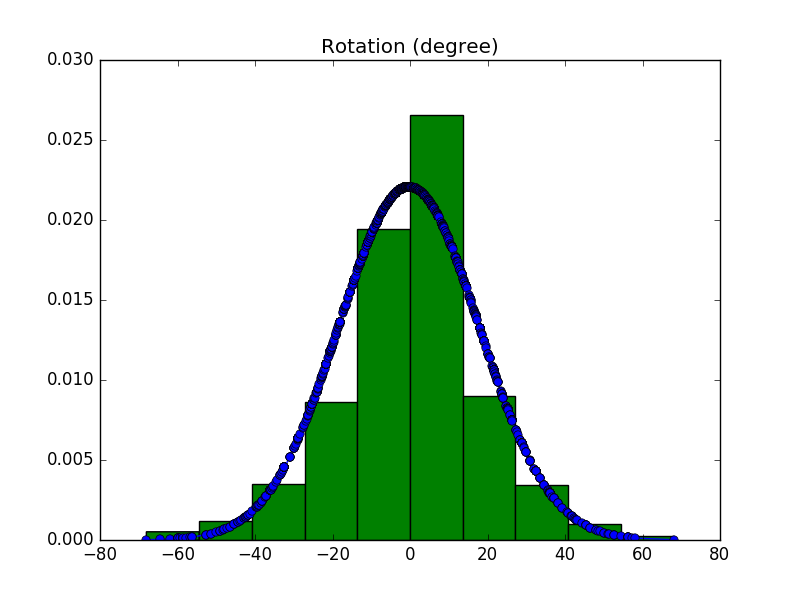}}} \hspace{1em}%
\subcaptionbox{\centering Growth (pixel)}{\fbox{\includegraphics[width=6cm]{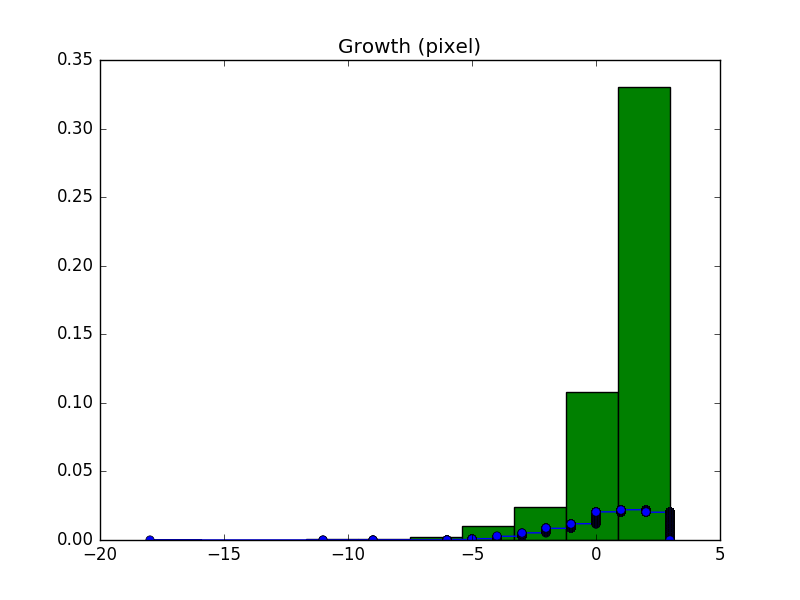}}}
\caption{Distributions according to our fit models according to accepted universes}
\label{fig:distributions}
\end{figure}

Figure \ref{fig:distributions} plots histograms of how much individual bacteria move, rotate, and lengthen between video frames (that is, these are ``delta'' changes, per-frame). We can see that all the movement and rotation distributions seem to fit a Gaussian curve, while growth in length is more constant at about 2 pixels per frame. We allow slight negative growth not because it is feasible in real life, but because it allows the simulation to correct minor over-estimates in length that can occasionally accumulate over a few consecutive frames.

We note that currently, the simulation code {\em chooses} values of these deltas from a uniform distribution, but then universes are selected (essentially in a Darwinian fashion) based upon actual movement observed. This means that many proposed movements in the simulation will be discarded as being too far from the distributions shown in Figure \ref{fig:distributions}. If we were instead to use the observed distributions, the simulation would become significantly more efficient (requiring less CPU) because fewer universes would be discarded as useless. In fact gathering these distributions dynamically {\em during} the simulation will allow us to {\em automatically} improve the efficiency of the simulation as it progresses; this is one area of future work.

\section{Discussion and future work}

\begin{figure}[htp]
\centering
\subcaptionbox{\centering ``Noisy" Frame 67}
{\fbox{\includegraphics[width=6cm]{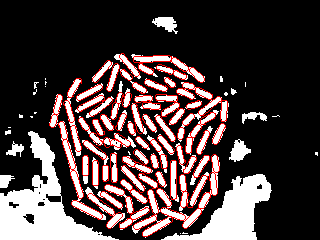}}} \hspace{1em}%
\caption{The individual bacteria along the edges of the video grow long enough that our software sometimes overcounts them.}
\label{fig:long-guys}
\end{figure}

As mentioned in the discussion around Figure \ref{fig:counts}, our method tends to overcount the number of bacteria present in later frames of the video. This occurs at least in part because the rules of our simulation do not take into account the possibility that bacteria may grow to be very long before splitting. Consider the long bacterium along the left edge of Frame 67 (Figure \ref{fig:long-guys}). Although it shows no visible signs of splitting, our software counts it as three bacteria because of its length. In contrast with the counting errors seen in the center of the video (see Figure \ref{fig:1st-fuckup}), which stem from the low quality of the video and cannot be corrected, this over-counting is an issue (albeit an easily-remedied one) with the ``rules of the universe'' encoded in our simulation. The best way to fix this is to integrate a more sophisticated hypothesis testing system into the simulation; this should allow us to detect a very long bacterium when it has no obvious ``break'' points, rather than spuriously splitting it simply because we cannot model a long bacterium.

Our current code is not particularly efficient, but could be made so with many conceptually simple optimizations: converting from Python to C++; better hypothesis testing, which should drastically reduce the number of parallel universes required to accurately track reality; and using more accurate probability density functions from which to draw our random distributions. It should also be possible to automatically optimize most simulation parameters and distributions.

Finally, considering that there already exist excellent user-interfaces for cell tracking such as CellProfiler, it would be fruitless to invent another user interface for cell tracking. Thus we intend to implement our code as a plug-in to CellProfiler, providing it with more reliable and robust automation than it currently has.

\section{Conclusion}

Here we have presented a novel method for accurate and robust tracking of cells even in the presence of significant noise that completely overwhelms current methods. We use simple physical rules to limit possibilities, which greatly enhances our ability to discern noise from signal, producing highly specific and sensitive measures of cell motility and precise detection of events such as cell divisions. Even in its infancy, this method produces cell counts comparable or superior to existing mature software, and {\em greatly} outperforms them in the details of precision modeling.

\section{Supporting Information}

\subsection*{Videos}

\label{S1_Video}
{\bf YouTube.com video that we used as our example throughout this paper: {\tt https://www.youtube.com/watch?v=UccyM8QeIeE}.}

\nolinenumbers

%
%
%
\bibliographystyle{plos2015}
\bibliography{sample}

\end{document}